\documentclass[twocolumn,english,aps,superscriptaddress, pra,groupedaddress]{revtex4-1}

\usepackage[latin9]{inputenc}
\usepackage{amsmath}
\usepackage{amssymb}
\usepackage{graphicx}
\usepackage{babel}
\usepackage{mathrsfs}
\usepackage{amsfonts}
\usepackage{epstopdf}
\usepackage{caption}
\usepackage{multirow}

\numberwithin{equation}{section}

\begin{document}

\title{Universal Scaling Properties of Cold Atom Scattering Dynamics \\
	 in Confined Low Dimensional Geometries}

\author{Jeff Maki}
\author{Fei Zhou}

\affiliation{Department of Physics and Astronomy, University of British Columbia, Vancouver V6T 1Z1, Canada}

\date{Mar. 6, 2014}

\begin{abstract}

Intermediate energy scale physics plays a very important role in non-equilibrium dynamics of quasi-low dimensional cold atom systems. In this article we obtain the universal scaling relations for the generalized reflection coefficient, i.e., the fraction of atoms scattered out of the initial state, at intermediate energy scales, scales larger than the trap frequency but much smaller than the effective range of the potential, for one and two dimensional harmonically confined geometries. Whenever the energy of the cold atoms is commensurate with a transverse energy level, it is shown that the system becomes non-interacting.  When the difference between the energy of the cold atoms and the given transverse energy level, $\delta E$, is small compared to the trap frequency, $\omega$, i.e. when $\delta \mathcal{E} = \delta E / \omega \ll 1$, the reflection coefficient has the universal scaling form $R \approx C \sqrt{\delta \mathcal{E}}$, where $C$ is a constant. The power law behaviour and prefactor $C$ appear regardless of the three dimensional scattering length and initial conditions.

\end{abstract}

\maketitle

\section{Introduction}

The study of cold atoms has flourished due to the vast extent of control and tunability available to experimentalists. In the laboratory it is now possible to create non-equilibrium systems in a controllable manner for any dimension, using optical traps, and for any interaction strength with Feshbach Resonance  \cite{Bloch, Chin_1}. Such experiments have led to the study of quench dynamics and metastable states \cite{Hulet,Wieman, Jila, Greiner, Review, Trotzky, Langen}, thermalization rates \cite{Petrov, Kinoshita, Mazets, Review} and the dynamics of solitons \cite{Khaykovich, Strecker}. These systems offer a rich variety of new physics in relation to their equilibriated counterparts. One of the major challenges of the field of dynamics is the presence of different energy scales in the problem. Far-from-equilibrium physics requires the knowledge of numerous energy scales which can differ appreciable, in direct contrast to thermodynamic physics where only energy scales small compared to the temperature dominate. These additional energy scales complicate the problems appreciably as the eigenstates of macroscopic systems at these larger scales are extremely complex.

In the field of dynamics, an interesting issue is whether and under what conditions there are universal properties at these intermediate energy scales, analogous to the ones seen in thermodynamics. Since the number of dynamical effects in cold atoms is vast, such a classification of dynamics into different universality classes would be an extremely powerful tool. Some efforts in this area have led to universal relations for quench dynamics across quantum critical points \cite{Review, Kibble, Polkovnikov, Sachdev1}, turbulence, and soliton dynamics in cold atom systems \cite{Nowak}.

Universal or not, one particularly well studied dynamical effect is the scattering properties of cold atoms in confined geometries \cite{Petrov, Mazets, Olshanii, Olshanii2, Schmelcher, Haller, Chin, Saeidian}. These scattering experiments provide detailed information about the interaction between individual atoms, and can be used as a starting point for many-body theories. The scattering properties of cold atoms are generally examined at low energy scales, where the interaction can be parametrized by the d-dimensional effective scattering length \cite{Maki}. In the case of harmonically confined geometries, it is possible to integrate out the transverse degrees of freedom and write down an effective low-dimensional model with scattering lengths defined in terms of the harmonic trap length, $a_{\perp}$, and the three dimensional scattering length, $a$ \cite{Olshanii, Petrov}. Furthermore, this allows one to tune the low-dimensional interaction strength from zero to infinitely positive or negative values by simply varying the three dimensional scattering length or the length scale of the harmonic confining potential. This effect is known as confinement induced resonance \cite{Olshanii}, and is a consequence of virtual scattering between different transverse energy levels in harmonically confined geometries. This phenomenon has been experimentally verified \cite{Haller, Haller2, Olshanii2} for both 1D and 2D systems. 

In this article we examine the scattering properties of cold atoms and introduce another important consequence of harmonically confined geometries: the universal properties of scattering dynamics at intermediate energy scales, or energies larger than the trap frequency $\omega_{\perp}$ but much smaller than the effective range of the potential. In general the scattering dynamics of cold atoms are not universal in the sense that they rely on the specific value of the effective low dimensional interaction. However, whenever the energy of the cold atoms is resonant with that of a transverse energy level, the atoms become non-interacting. In the proximity of such a resonance, physical observables take on universal scaling forms depending only upon the energy difference between the atoms and the transverse energy level. The scaling forms in this limit are independent of the interaction strength and for a wide range of initial conditions.

\section{Two Body Problem for Arbitrary Low Dimensions}

Consider two cold atoms in the presence of a confining potential $U(\vec{r}) = \frac{1}{2} m \omega^2 r^2_{\perp}$, where $r_{\perp}$ represents the magnitude of the position vector in the transverse directions, interacting via a separable contact interaction $V(\vec{r}-\vec{r}') = g \delta(\vec{r}- \vec{r}')$. Such a confinement can be easily produced in the laboratory \cite{Bloch, Chin_1} and acts to reduce the original three dimensional system to that of a quasi-d dimensional system, where $d= 2-\epsilon$, and $\epsilon > 0$. In the case of harmonic confinement, the center of mass and relative motions of these two particles can be separated. The interaction will only affect the relative motion of the particles, and thus the center of mass motion can be ignored. The Hamiltonian for the relative motion is given as:

\begin{equation}
H = -\nabla^2 + U(\vec{r}) + V(\vec{r}),
\label{eq:Hamiltonian}
\end{equation}

\noindent where $m$ and $\hbar$ have been set to unity. The relative Hamiltonian without interactions has a set of eigenstates with continuous quantum numbers $\vec{k}$ for the free $d=2-\epsilon$ dimensions, and a discrete set, $n$, for the transverse energy levels. The energy for a state $\left| \nu \right\rangle = \left| \vec{k}, n \right\rangle$ is $E_{\nu} = k^2 + E_n$. The discrete energy levels associated with the transverse motion satisfy the harmonic oscillator Schrodinger equation in the transverse dimensions. In this notation, the spectrum of the Hamiltonian is a series of bands with quadratic dispersion indexed by $n$.

All the scattering dynamics are contained in the T-matrix which incorporates the effects of the interaction to infinite order. The T-matrix is defined through the integral equation:

\begin{equation}
\left\langle \nu \left| \hat{T} \right| \nu' \right\rangle = \left\langle \nu \left| \hat{V} \right| \nu' \right\rangle + \sum_{\nu''} \left\langle \nu \left| \hat{V}\right| \nu'' \right\rangle G_{\nu''} \left\langle \nu'' \left| \hat{T} \right| \nu' \right\rangle,
\label{eq:Tmatrixdef}
\end{equation}

\noindent where $G_{\nu}$ is the propagator for the non-interacting Hamiltonian. Since the interaction potential is separable, it is possible to solve the scattering problem in its entirety. The solution is:

\begin{equation}
T(E)^{-1} = \left[\frac{1}{g} - \frac{\Gamma\left(\epsilon / 2\right)}{(4 \pi)^{1 - \epsilon / 2}} \sum_n \frac{\left| \phi_n(0) \right|^2}{\left( E_n - E - i \delta \right)^{\epsilon / 2}} \right].
\label{eq:general_T}
\end{equation}

\noindent In Eq.~(\ref{eq:general_T}), the free dimensions have been integrated out, leaving only the discrete bands associated with the transverse motion. The sum in Eq.~(\ref{eq:general_T}) is divergent. This divergence can be incorporated into a finite renormalized coupling constant $g$. Each band contributes to the renormalization process, and it is possible to study the flow of the coupling constant as these transverse energy levels are removed from the system.

To this end consider a system initially prepared in the state $\left| \nu_0 \right\rangle = \left| n_0, \vec{k} \right\rangle$ with energy $E$. Instead of a system with an infinite number of bands, one considers a system with a finite number of bands with indices $n_0 -s \leq n \leq n_0 + s$ and an appropriate renormalized coupling constant:

\begin{equation}
T(E)^{-1} = \left[\frac{1}{g(s)} - \frac{\Gamma\left(\epsilon / 2\right)}{(4 \pi)^{1 - \epsilon / 2}} \sum_{n=n_0-s}^{n_0+s} \frac{\left| \phi_n(0) \right|^2}{\left( E_n - E - i \delta \right)^{\epsilon / 2}} \right].
\label{eq:T(s)}
\end{equation}

\noindent The renormalized coupling constant, $g(s)$, is a function of the initial conditions,  the energy and, the number of bands left in the system, parametrized by $s$; $g(s, E, n_0) \equiv g(s)$. We require that the T-matrix produces the same physics regardless of the number of bands, which implies that the T-matrix is invariant for any value of $s$. This invariance condition allows us to calculate the change in the coupling constant as $s$ is reduced to zero. The renormalization step is defined by integrating out the bands $n_0+s$ and $n_0-s$ (if it exists) by defining a new renormalized coupling constant $g(s-1)$:

\begin{eqnarray}
\frac{1}{g(s-1)} - \frac{1}{g(s)} &=&  \frac{\Gamma(\epsilon / 2)}{(4 \pi)^{1 - \epsilon / 2}} \left( \frac{\left| \psi_{n_0+s}(0) \right|^2}{\left( E_{n_0+s} - E - i \delta \right)^{\epsilon / 2}} \right. \nonumber \\
&& \left. + \frac{\left| \psi_{n_0-s}(0) \right|^2}{\left( E_{n_0-s} - E - i \delta \right)^{\epsilon / 2}} \right),
\end{eqnarray}

\begin{figure}
\includegraphics[angle = 270, scale = 0.35]{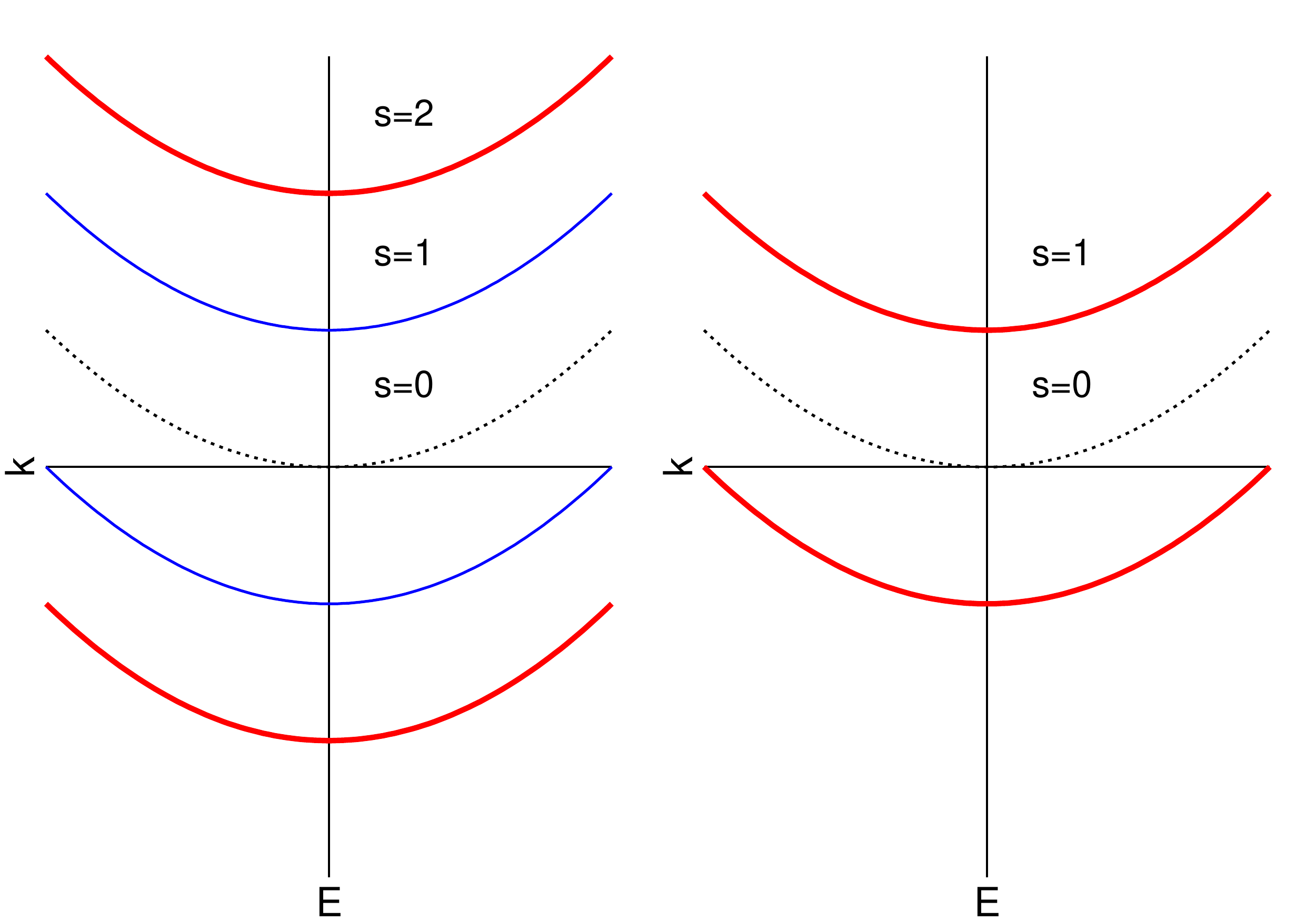}
\caption{(Color online) The above figure shows a single step in the renormalization procedure. Before the step the system is characterized by a finite number of bands with indices $n_0 - s \leq n \leq n_0+s$ and a coupling constant $g(s)$, which is shown by the left hand diagram. The bands with indices $n_0 \pm s$, designated by the thick red lines, are then simultaneously integrated out. The result is a system with two less bands and a new coupling constant $g(s-1)$ as shown on the right.}
\label{fig1}
\end{figure}

\noindent This procedure is performed until the system only contains one band left, the initial one, and is shown in Fig.~(\ref{fig1}).

After performing the renormalization step, one finds the relation between $g(s)$ and $g(s-1)$:

\begin{eqnarray}
\Delta g(s) &=& g(s) - g(s-1) \nonumber \\
&=& \frac{1}{4 \pi a_{\perp}} g(s)^2 I(s) \left( 1 + \frac{g(s) I(s)}{4 \pi a_{\perp}} \right)^{-1}.
\label{eq:delta_g}
\end{eqnarray}

\noindent where $I(s)$ is:

\begin{eqnarray}
I(s) &=& 4 \pi a_{\perp} \frac{\Gamma(\epsilon / 2)}{(4 \pi)^{1 - \epsilon / 2}} \left( \frac{\left| \psi_{n_0+s}(0) \right|^2}{\left( E_{n_0+s} - E - i \delta \right)^{\epsilon / 2}} \right. \nonumber \\
&& \left. + \frac{\left| \psi_{n_0-s}(0) \right|^2}{\left( E_{n_0-s} - E - i \delta \right)^{\epsilon / 2}} \right),
\label{eq:I}
\end{eqnarray}

\noindent and $a_{\perp}$ is the harmonic length of the trap. It is then possible to cast this relation into a form that resembles the standard renormalization group formulation \cite{Sachdev, PandS}. One first defines a dimensionless coupling constant,

\begin{equation}
\tilde{g}(s) = \sqrt{2 \omega (s-1)} g(s) = \frac{2}{a_{\perp}} \sqrt{s-1} g(s),
\end{equation}
\noindent  With this definition, in direct analogy to the $\beta$-function, we define $\alpha(\tilde{g},s,E)$, the so called $\alpha$-function, to describe the flow of $\tilde{g}$: 

\begin{eqnarray}
\alpha(\tilde{g},s,E) &=& \frac{\Delta{\tilde{g}(s)}}{\Delta \log(s-1)} \nonumber \\&=& \left(s-1 - \sqrt{(s-1)(s-2)} \right) \tilde{g}(s) \nonumber \\
&+& \frac{1}{8 \pi}\sqrt{s-2} \frac{\tilde{g}^2(s) I(s)}{1 + \frac{1}{8 \pi} \frac{\tilde{g}(s) I(s)}{\sqrt{s-1}}}.
\label{eq:alpha}
\end{eqnarray}

In contrast to the $\beta$-function, the $\alpha$-function depends explicitly on the cut-off $s$. This is due to the fact that the harmonic trap explicitly breaks the scale invariance of the problem. In spite of this, the $\alpha$-function still has relevant information on the renormalization process in such a system. One can show that the only fixed point in Eq.~(\ref{eq:alpha}) is the non-interacting fixed point, $\tilde{g}(s) =0$, opposed to the $\beta$-function which allows a strong coupling fixed point in the infra-red limit in the case of one dimension, the Tonks-Girardeau gas \cite{TG}. This non-interacting fixed point can be reached in one of two ways. The first is the trivial case where the system is initially non-interacting and $\alpha = 0$ for all steps. However, it is possible to reach this fixed point in an atypical manner. 

\begin{figure*}
\begin{center}
\includegraphics[width=0.8\textwidth, height = 8cm]{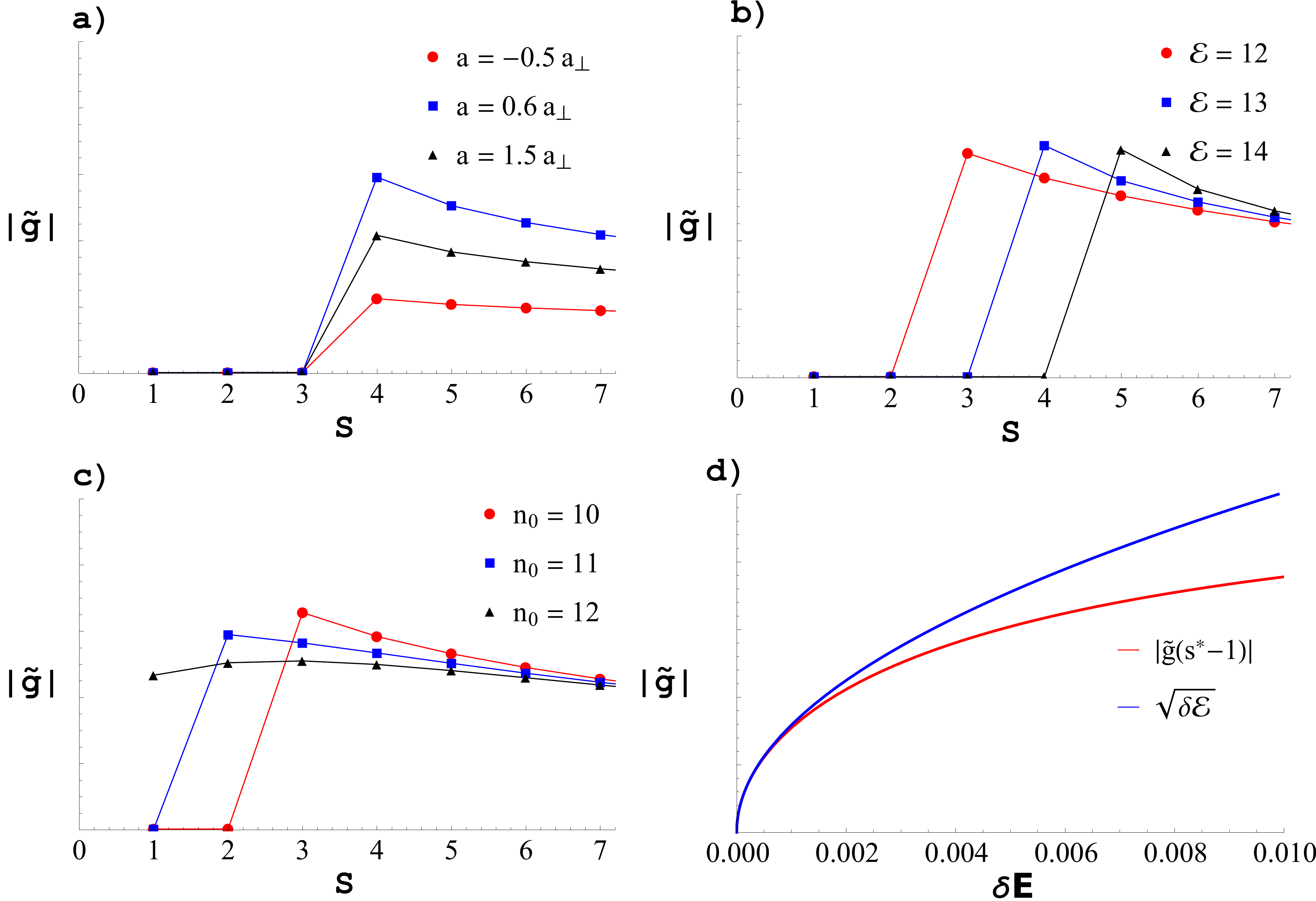}
\caption{(Color online) The discrete flow of the dimensionless coupling constant, $\tilde{g}$, is shown above under three different conditions: {\bf a)} fixed initial band $n_0$, and  total energy $\mathcal{E}$, with varying interaction strength $a$ {\bf b)} fixed $n_0$ and $a$ with varying $\mathcal{E}$, and {\bf c)} fixed $\mathcal{E}$ and $a$ with varying $n_0$. In {\bf a)} the system is prepared with $n_0 = 10$, and $\mathcal{E} \approx 13$. The dimensionless coupling constant will universally flow into the non-interacting fixed point when $s=3$ regardless of $a$.  A similar situation is shown in {\bf b)}. For the same $n_0$ and $a$, the initial energy is varied in units of the band gap. $\tilde{g}$ once again universally renormalizes to zero at the step $s$ with $\mathcal{E} = n_0+s$. Finally, in plot {\bf c)} $a$ and the energy $\mathcal{E}$ are fixed with $\mathcal{E} = 12$. The flow is shown for three different choices of the initial band index $n_0$. $\tilde{g}$ then renormalizes to zero for all $n^* \neq n_0$. Only in the case $n^* = n_0$ will $\tilde{g}$ renormalize to some non-zero value dictated by the true low dimensional system. In part {\bf d)} the value of the coupling constant after renormalization $|\tilde{g}(s^*-1)|$ is shown for different values of $\delta \mathcal{E}$ with $n_0 = 10$, $s = 2$, and $\mathcal{E} = 12 + \delta \mathcal{E}$ for a quasi-one dimensional system. For finite $ \delta \mathcal{E}$, there are corrections to the universal form of $| \tilde{g}(s^*-1)| \propto \sqrt{\delta \mathcal{E}}$ of order $\delta \mathcal{E}$.}
\label{fig2}
\end{center}
\end{figure*}

If the initial dimensionless coupling constant is non-zero, it is possible to reach the non-interacting case if $I(s)$ is divergent for some step $s^*$. At this step the $\alpha$-function has the following form $\alpha(g,s^*,E) = (s-1) \tilde{g}(s^*)$, which implies $\tilde{g}(s^*-1) = 0$, independent of the actual form of $\tilde{g}(s^*)$ before the step. This universality of the flow is shown in Fig.~\ref{fig2} (a). The fixed point is thus universal for infinite $I(s^*)$. If $I(s^*)$ is finite, non-universal corrections that depend on $\tilde{g}(s^*)$ will appear and define the width of the universal regime.

For the specific case of Eq.~(\ref{eq:I}), $I(s^*)$ diverges whenever the energy is commensurate with $E_{n^*}$ as $\epsilon^{-1}\left(\delta E \right)^{-\epsilon / 2}$, where $\delta E = \left| E - E_{n*} \right|$ and $n^* = n_0 + s^*$. This creates a sequence of non-interacting universal fixed points separated by the band gap in energy space as demonstrated Fig.~\ref{fig2} (b) and (c). For large but finite $I(s^*)$, the dimensionless coupling constant will have the universal scaling form:

\begin{equation}
\tilde{g}(s^*-1) \propto \delta E_{n^*}^{\epsilon /2} / \epsilon. 
\end{equation}

\noindent Corrections to this universal behaviour will appear at order $\delta E_{n^*}$ and are depicted in Fig~\ref{fig2} (d). In the limit $\epsilon=0$, or in two dimensions, $I(s^*)$ diverges logarithmically and hence the dimensionless coupling constant will also scale logarithmically, $\log^{-1} |\delta \mathcal{E}|$, with corrections of order $\log^{-2} | \delta \mathcal{E}|$. 

However, if the kinetic energy of the system is much less than the band gap, the initial band and $n^*$ will coincide. Thus the interaction will renormalize to a finite value as the initial band is not included in the renormalization procedure. The system then behaves as a true low dimensional system which, for one dimensional gases, implies the system will form a Tonks-Girardeau gas, whereas in two dimensions, the gas will instead evolve towards the non-interacting gas. 

This unique structure of the $\alpha$-funciton has immediate implications on the scattering observables of quasi-low dimensional gases. This can be deduced by examining the fraction of particles scattered out of an initial state, i.e. a generalized reflection coefficient. Near these non-interacting fixed points we can calculate the reflection coefficient perturbatively by normalizing the transition rate from Fermi's golden rule \cite{Sakurai} with respect to the initial particle flux. The matrix element, $\langle f | V | i \rangle$ can be replaced with the renormalized coupling constant as it incorporates the effect of all the possible interactions. Following the above prescription for the reflection coefficient one finds:

\begin{equation}
R = C \frac{\eta( \mathcal{E})}{\sqrt{\mathcal{E} - n_0}} |\tilde{g}|^2.
\label{R_prediction}
\end{equation}

\noindent where $\mathcal{E}$ is the energy measured in units of the band gap, which is $2 \omega$ for quasi-one and quasi-two dimensional systems, and $\eta(\mathcal{E})$ is the density of states. The constant $C$ is universal and depends specifically on the flow of the coupling constant.

Slightly above a transverse energy level $\eta(\mathcal{E})$ diverges like $\delta \mathcal{E}^{-\epsilon/2}$, with $\delta \mathcal{E} = \mathcal{E} - n$, but is otherwise finite. When the energy is commensurate with a harmonic oscillator level, Eq.~(\ref{R_prediction}) will exhibit a universal scaling form. For $\delta \mathcal{E} > 0$, a partial cancellation occurs between the renormalized coupling constant and the density of states leading to the universal scaling relation for $R$:

\begin{equation}
R = C \delta \mathcal{E}^{\epsilon / 2} / \epsilon.
\end{equation}

\noindent When $\delta \mathcal{E} < 0$, the density of states is finite and as a result there is a different scaling relation:

\begin{equation}
R = C \delta \mathcal{E}^{\epsilon} / \epsilon.
\end{equation}  

\noindent Finally, when $\epsilon = 0$. The case of a quasi-two dimensional systems, the density of states is a constant, and there is only one scaling relation for $R$:

\begin{equation}
R = C \log^{-2}|\delta \mathcal{E}|,
\end{equation}

\noindent valid whenever $| \delta \mathcal{E} | \ll 1$.  The scaling exponents for these relations are found in Table~\ref{tab:1} for quasi-one and quasi-two dimensional systems.

\begin{table}
\begin{center}
\begin{tabular}{|c|c|c|}
\hline
\multirow{2}{*}{Dimension} & \multicolumn{2}{|c|}{Scaling Exponent} \\
\cline{2-3}
 & $ \delta \mathcal{E} < 0$ & $ \delta \mathcal{E} > 0$ \\
\hline
1D & $0.5$ & $1$\\
\hline
2D & $2$ & $2$ \\
\hline
\end{tabular}
\end{center}
\caption{Universal scaling exponents for quasi-one and quasi-two dimensional systems when the energy, in units of the transverse energy scale, $\mathcal{E}$, is commensurate with a transverse energy level. In the case of 1D systems, the scattering observables scale as $\left| \delta \mathcal{E} \right|^x$, for some exponent $x$. Similarly for 2D, the observables scale as $\log^{-x} \left| \delta \mathcal{E} \right|$. The scaling coefficients are shown for the two possibilities; when the energy approaches some excited band from below or above.}
\label{tab:1}
\end{table}

We now compare Eq.~(\ref{R_prediction}) with the exact results obtained for quasi-one and quasi-two dimensional systems. For quasi-one dimensional gases, a calculation of the reflection coefficient for arbitrary energies is given by \cite{Olshanii}:

\begin{equation}
R = \eta(\mathcal{E}) \left| \frac{a_{\perp}}{a} + \zeta(1/2, -\mathcal{E}) \right|^{-1},
\label{R_exact}
\end{equation}

\noindent where $\zeta(1/2,x)$, is the Hurwitz zeta function \cite{Stegun}.

For positive $\mathcal{E}$, the Hurwitz zeta function diverges periodically whenever $\mathcal{E}$ is equal to an integer, or equivalently, when the energy is near to a harmonic oscillator level, as $\delta \mathcal{E}^{-1/2}$. Upon expanding this exact solution around the non-interacting fixed points one finds the same universal behaviour and scaling exponents as the renormalization analysis.

Our perturbative analysis of the reflection coefficient holds for every non-interacting fixed point. When $n^* = n_0$ the system will renormalize into a strong coupling regime. In this case the exact solution gives:

\begin{equation}
1-R \approx \delta \mathcal{E} \left( \frac{a_{\perp}}{a} + \zeta(1/2) \right),
\label{eq:R_zero}
\end{equation}

\noindent fully consistent with the predictions from the $\alpha$-function. 

For a quasi-two-dimensional system, it is the cross section $\sigma$ that exhibits periodic divergences whenever $\mathcal{E}$ is an integer. The exact solution for $\sigma$ is found in Ref.~\cite{Petrov}. This solution has the same scaling behaviour as found by Eq.~(\ref{R_prediction}). The flow for quasi-two and two dimensional systems are both described by non-interacting fixed points.

\section{Conclusion}

In this analysis we considered a discrete renormalization procedure to study the contact interaction between two cold atoms in a confined low dimensional geometry. The renormalization procedure was performed by integrating out each band describing a transverse energy level. If the energy of the system was commensurate with the bottom of a new band ($n ^*\neq n_0$), the interaction then renormalized to zero, resulting in total transmission and a non-interacting theory. Near these critical points, observables take on universal scaling forms and their scaling exponents were determined. This approach is valid for a wide range of initial conditions, and as a result is quite ubiquitous. The only caveat is when the kinetic energy of the initial system is much less than the band gap $2 \omega$, i.e $n^*  = n_0$. In this case the interaction renormalizes to a constant, and the physics is dominated by the true low dimensional system. For quasi-one dimensional systems, this observation leads to the strong coupling Tonks-Girardeau gas \cite{TG}. 

The question of universal dynamics is still in its infancy, but has resulted in several fascinating results in quench dynamics \cite{Review,Kibble, Polkovnikov, Sachdev1} and cold atom systems \cite{Nowak}. It remains to be seen whether other dynamical systems can fall into universality classes. The number of different dynamical effects is quite large, and the conditions under which universal dynamics can be obtained is a pertinent question. If an understanding of when universal dynamical effects are present, the categorization of different universality classes can be conducted thoroughly and explicitly tested in the laboratory. 

The authors would like to thank Shao-Jian Jiang for useful discussions. This work was funded by the Canadian Institute for Advanced Research (CIFAR) and the Natural Sciences and Engineering Research Council of Canada (NSERC). 

In preparing this manuscript, a similar work was done in Ref.~\cite{Heb} using numerical solutions of an exact K-matrix formalism.

\end{document}